\documentclass[journal]{IEEEtran}

\usepackage{vcell}
\usepackage{rotating}
\usepackage{xcolor,soul,framed} %,caption

\colorlet{shadecolor}{yellow}
\graphicspath{{../pdf/}{../jpeg/}}
\DeclareGraphicsExtensions{.pdf,.jpeg,.png}
\usepackage{colortbl}
\usepackage{multirow}
\usepackage{makecell}
\usepackage{hhline}
\usepackage{graphicx}
\usepackage[cmex10]{amsmath}
\usepackage{array}
\usepackage{mdwmath}
\usepackage{mdwtab}
\usepackage{eqparbox}
\usepackage{url}
\usepackage{amssymb}
\usepackage{caption}
\usepackage{subfigure}
\usepackage{siunitx}

\begin{document}
\bstctlcite{IEEEexample:BSTcontrol}
    \title{Stereo X-ray Tomography}
  \author{Zhenduo~Shang and
      Thomas~Blumensath% <-this % stops a space

  \thanks{Pedro Galvez Hernandez provides the real data, https://research-information.bris.ac.uk/en/persons/pedro-galvez-hernandez from the Universty of Bristol }
  \thanks{Nvidia Graphic cards provides computation resource on deep learning model.}}% <-this % stops a space
%   \thanks{XXXXX.}}  

\markboth{
}{Roberg \MakeLowercase{\textit{et al.}}: XXXXX}

% ====================================================================
\maketitle

% === ABSTRACT ====================================================================
% =================================================================================
\begin{abstract}
%\boldmath
X-ray tomography is a powerful volumetric imaging technique, but detailed three dimensional (3D) imaging requires the acquisition of a large number of individual X-ray images, which is time consuming. For applications where spatial information needs to be collected quickly, for example, when studying dynamic processes, standard X-ray tomography is therefore not applicable. Inspired by stereo vision, in this paper, we develop X-ray imaging methods that work with two X-ray projection images. In this setting, without the use of additional strong prior information, we no longer have enough information to fully recover the 3D tomographic images. However, up to a point, we are nevertheless able to extract spatial locations of point and line features. From stereo vision, it is well known that, for a known imaging geometry, once the same point is identified in two images taken from different directions, then the point's location in 3D space is exactly specified. The challenge is the matching of points between images. As X-ray transmission images are fundamentally different from the surface reflection images used in standard computer vision, we here develop a different feature identification and matching approach. In fact, once point like features are identified, if there are limited points in the image, then they can often be matched exactly. In fact, by utilising a third observation from an appropriate direction, matching becomes unique. Once matched, point locations in 3D space are easily computed using geometric considerations. Linear features, with clear end points, can be located using a similar approach. 

%%% %%% %%% %%% %%% %%% %%% %%% %%% 
% Maybe move that into introduction, does not sit well in abstract. Instead, include highlight of results.
%%% %%% %%% %%% %%% %%% %%% %%% %%% 
% Our approach takes a different approach from work that builds full tomographic reconstructions from limited observations. Whilst our work is only able to locate point and line features, we do not rely on strong prior image models, which would be necessary for full tomographic reconstruction.

\end{abstract}

% === KEYWORDS ====================================================================
% =================================================================================

% \fi
\IEEEpeerreviewmaketitle

% ====================================================================
% ====================================================================
% ====================================================================

% === I. INTRODUCTION =============================================================
% =================================================================================
\section{Introduction}

\IEEEPARstart{X}{R}ay Computed Tomography (XCT) is a powerful volumetric imaging technique, widely applied in medical and industrial applications to reveal the internal structure of an object by collecting thousands of projection images from different views around the object. The achievable image quality depends on the number of acquisitions. With the help of advanced algorithms and using regularisation functions such as Total Variation (TV) that enforce certain image smoothness properties \cite{song2007sparseness}, or learned, data-driven regularises \cite{yuan2018sipid}, \cite{lee2018deep}, \cite{pelt2022cycloidal}, \cite{ernst2021sinogram}, \cite{adler2018learned}, a significant reduction in the number of measurements is possible, without a significant reduction in image quality. In fact, a range of recent papers has shown that even single projection images are sufficient in certain settings to identify one known image out of a small selection of possible reconstructions (see for example \cite{li20213}) though this only works by imposing very strong prior constraints on the reconstruction, which is only possible if the objects are extremely predictable in their 3D shape.

\begin{figure}
  \begin{center}
  \includegraphics[width=3.5in]{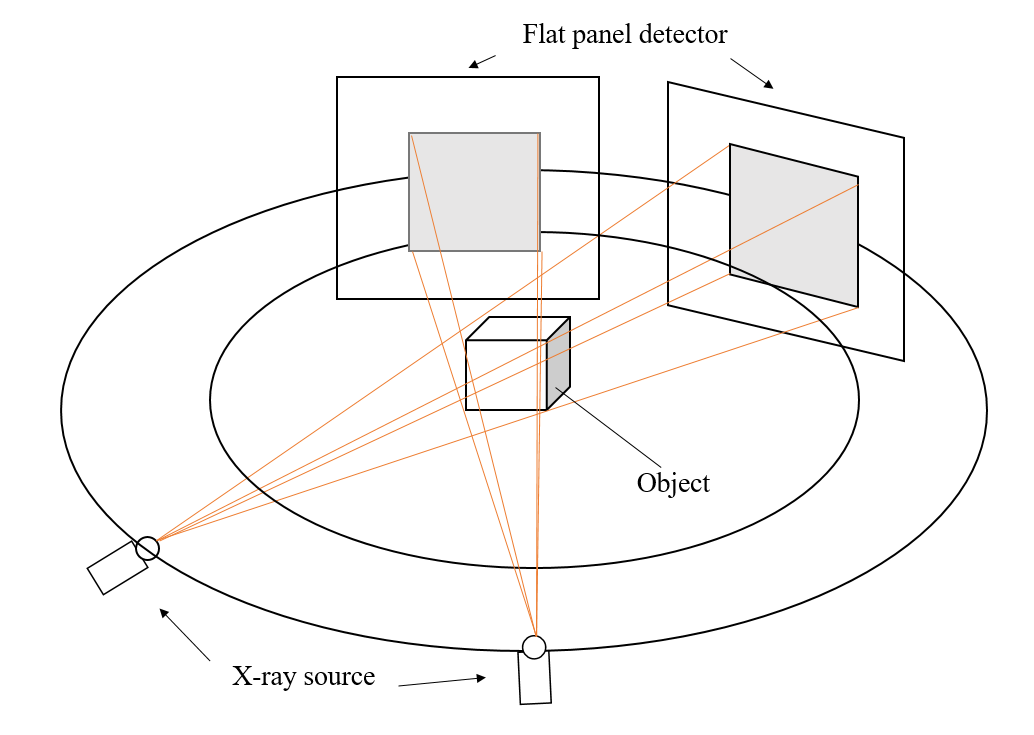}\\
  \caption{For stereo X-ray tomographic imaging with two views, two X-ray projection images are taken from an object from two different directions.}\label{Two X-ray Sources Schematic}
  \end{center}
\end{figure}

Our work takes a different approach. Instead of trying to reconstruct full tomographic images from limited observations, which require very strong prior information, we only aim at the recovery of the 3D location of point and line like features. We are here interested in time-sensitive applications, where the internal structure of an object changes rapidly, and where we might only be able to take a single image at each time-step. Inspired by computer stereo vision systems, we assume an imaging setup with two or three X-ray source/detector pairs as shown in Fig. \ref{Two X-ray Sources Schematic}.

This allows us to take two or three projection images at the same time, but, without significantly restricting the volumetric images we are inspecting, this will not allow us to compute a full volumetric image as the two or three projection images simply do not contain enough information. Instead, we propose to only identify and locate specific internal features. In fact, we show that we are able to identify point and line like features and estimate their location in 3D space. To achieve this, we develop two deep neural networks, one that detects relevant features in the individual 2D projection images and one that matches these features and places them into 3D space. Whilst this will not provide the same level of information as full tomographic imaging, in several applications, this technique can nevertheless provide valuable dynamic information that is not accessible with traditional methods.

% =======

Our work is inspired by stereo vision applications, where two visible-light cameras are used to infer the location of visible objects in 3D space. With visible light, for non-transparent object, the amount of light measured at any point on a camera's image plane is assumed to be due to the reflection of light from a single point in the 3D scene. Thus, if we take two images of a scene from slightly different directions and if we are able to match points that come from the same point in the 3D scene between the two images, then a depth map for each image can be computed from simple geometric principles, placing each point in each image to a unique point in 3D space. The main issue in stereo vision is thus one of matching points between the two images of a stereo pair \cite{horn1986robot}.
% Not sure if this paper is relevant here, as this paper uses a single image, not stereo. This gets us too far from what we are talking about and has similar issues with single view tomography.
%\cite{eigen2014depth} demonstrate the method can generate the depth map from two 2D images by matching the point in space on two photos, 
% Not sure if this is relevant. Instead, can you cite a textbook here that describes the basics of stereo vision?
%Recent methods such as \cite{wu20153d} and \cite{jimenez2016unsupervised} infer the complete 3D voxel distribution from two 2D images using modern machine learning tools. 

Unlike stereo vision with visible light, in X-ray imaging, individual points on the imaging plane represent X-ray attenuation values along the entire X-ray path, that is, along the straight line from the X-ray source to the pixel on the X-ray detector \cite{kak2001principles}.  
% Put a citation here for basic X-ray CT, maybe the book by Kak and Slaney
Therefore, individual points on one imaging plane no longer correspond to a single point in 3D space, but to an entire line. For a full tomographic reconstruction, point matching based approaches are thus not possible. However, if there are point features in 3D space, such as the end of a linear feature, the corner of an object or an object that is of a similar size to the resolution of the system, then, this feature can be mapped in 3D space even from as few as two tomographic projections. To achieve this we need to 1) identify all point like features in the two X-ray views and 2) match these features between the two views. Once found and matched, mapping the features into 3D space uses the same geometric considerations as stereo vision. This approach can also be extended to line like features, by treating each line as a string of point like features, though here, matching can become more difficult.

% I think this can go into the methods section, not sure if you want to combine this with your METHODS section or split into two sections.
\subsection{Proposed method}

\begin{figure}
  \begin{center}
  \includegraphics[width=3.5in]{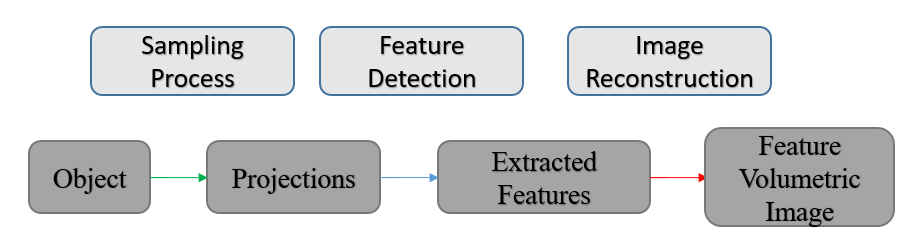}\\
  \caption{The proposed feature reconstruction framework. The sampling process, shown with the green arrow, represents the imaging process which generates a few projection images. Our approach then implements a feature detection step (blue arrow) and, once features are detected, a 3D mapping step (red arrow).}\label{Extracted features reconstruction frameworks}
  \end{center}
\end{figure}

Whilst there are many existing feature detection methods \cite{bay2006surf,bay2008speeded,lowe2004sift,viswanathan2009features,derpanis2004harris}, these have generally been developed for standard imaging applications and we found they do not work well on X-ray transmission images. In this paper, we thus propose a novel, deep learning based feature detector that can be trained on specific point and line like features. The particular challenge here is for the method to ignore the other image information that clutters X-ray projection images, but that is not related to point and line like features. The proposed framework is shown in Fig. \ref{Extracted features reconstruction frameworks}. After extracting the point and line like features in the 2D X-ray projection images, we then match the features by using the filtered back-projected (FBP) \cite{sagara2010abdominal}, \cite{hoffman1979quantitation} method to generate a 3D volumetric image of the extracted features. The back-projection of the features from the two images that are due to the same point in space will overlap in 3D space. If there are few features and if these features are randomly located in space, then it is likely that this intersection is unique, which will then lead to a unique match, though the more features we have, the more likely will it be that there will be an intersection of the lines in 3D space from more than one point in each image. In this case, an exact match is not possible from two observations, though by adding a third observation, this issue can be overcome. Another issue that can arise if there are many features in 3D space is that two features are aligned such that they both are measured in a single point on one of the imaging plains. To increase the robustness to the exact localisation of features on the two imaging planes, instead of simply looking at intersections in 3D space after back-projection, we use a simple deep neural network to process the back-projected volumetric image to generate a 3D image containing the point like features.

% =======

\subsection{Contributions}
The main contributions of this paper are:

    1) We design a convolutional neural network to identify point and line features in individual X-ray projection images that work even if the images show additional complex object features. %, the method is not only applicable to X-ray imaging, but also to the field of visible-light imaging.
    
    2) We design a deep neural network that uses pairs of feature images to map the features to their spatial location in 3D. This network in effect matches the features and uses geometric information to compute spatial locations.
    
    3) We evaluate the methods on two datasets, a synthetic dataset and a real X-ray imaging data-set.
% === II. PROPOSED APPROACH ========================
% =================================================================================
\section{Methodology}

\subsection{Feature Detection}
 There are many feature detection methods that have been developed for 2D images and common methods include the Speeded Up Robust Feature (SURF) method \cite{bay2006surf}, \cite{bay2008speeded}, Scale Invariant Feature Transform (SIFT) method \cite{lowe2004sift}, Features from Accelerated Segment Test (FAST) method \cite{viswanathan2009features} and Harris corner detection method\cite{derpanis2004harris}. However, these methods search for features commonly found in photographic images. They have not been designed to detect point and line like features in tomographic data and are thus not suitable for our task. Not only are they often unable to detect point and line features, they are also prone to detect other image structure in the tomographic background, which is not of interest here. Therefore, we design a binary classification neural network to detect features of interest. The feature detection problem can be summarized as finding parameters $\theta$ such that 
 \begin{equation}
  \mathbf{y}_{mask}  =  f\left ( \mathbf{x}_{raw};\theta \right )\label{eq:1.1}
\end{equation}
is a mapping from a raw 2D X-ray projection images $\mathbf{x}_{raw}\in \mathbb{R}^{m\times n}$ to a binary feature mask $\mathbf{y}_{mask}\in \mathbb{B}^{m\times n}$. The parameters are learned from training a dataset $\left \{ \left ( \mathbf{x}^{i}_{raw},\mathbf{y}^{i}_{mask} \right ) \right \}^N_{i=1}$ which comprises $N$ sample pairs. %Taking a raw X-ray projection $x$ as input, the output via the deep neural network is an estimation on extracted feature mask $\widehat{y}_{mask}=f\left ( \mathbf{x}_{raw};\Theta  \right )$. 

%%% This is not really required, you can work with the mask directly from here onwards.
%The estimation is then dot-multiplied with its corresponding $x_{raw}$ to get the extracted feature projection we need, the process is formulated as:

%\begin{equation}
    %\widehat{y}_{feature}=\widehat{y}_{mask}\cdot %x_{raw}\label{eq:1.2}
%\end{equation}

\subsection{Mapping Features into 3D space }

Once we have identified the features in the 2D projections, we could try and match points in projection pairs in the same way in which this is done in stereo vision. This is a two stage process with the difficult step being the matching of points in one projection to those in the other projection. In stereo vision, this matching is typically done by looking at the similarity of the image around a point and matching the points if their neighbourhoods are similar. This matching is further constrained by the fact that a point in one image can only match with points along a line in the other image, a constraint known as the epipolar constraint of the stereo camera model. In fact, if there are few points that need to be matched between two images, this constraint often means that there is a unique match as long as there is no more than one point feature on each epipolar plane in space. This process can thus also be used for matching points in our setting. Note that, for a trinocular X-ray stereo system, that is, for a system with 3 X-ray sources and detectors, where none of the epipolar planes of each of the three detector pairs is parallel, unique matching becomes increasingly likely, as now, not only do matching points lie on one epipolar plane, but on three epipolar planes \cite{trinocular}. To match linear features, we can start by matching the endpoints of a linear feature and then track the line from its endpoints. Once matched, point features can then be mapped into 3D space using knowledge of the camera geometry. 

\begin{figure}
  \begin{center}
  \includegraphics[width=3.5in]{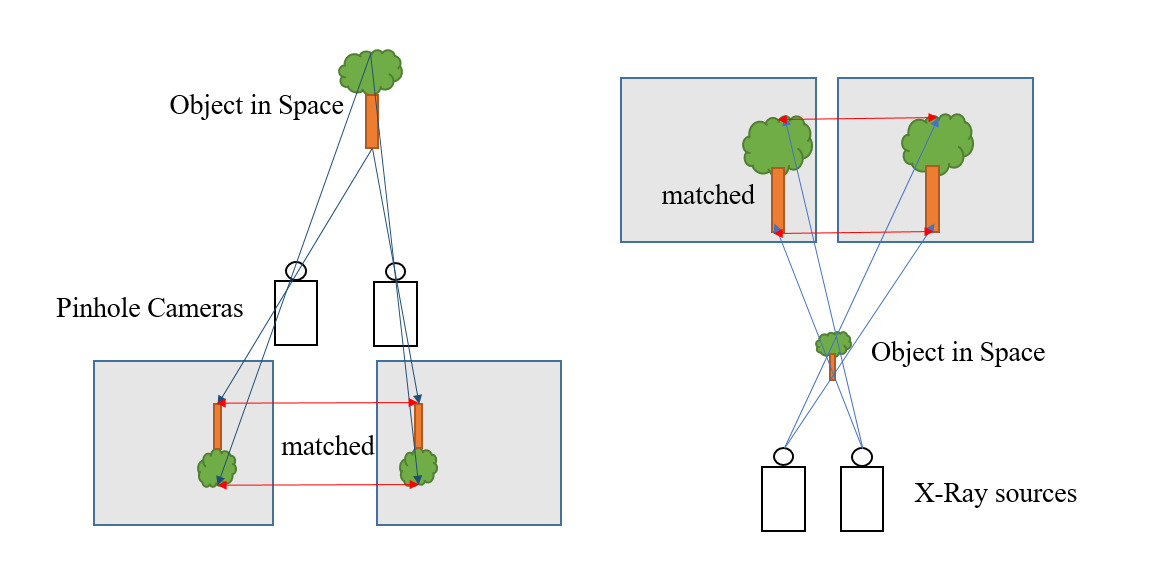}\\
  \caption{The geometry of visible-light stereo matching of point features using a pinhole camera model and X-ray projection point matching. }\label{point matching principles}
  \end{center}
\end{figure}

The similarity of the geometric depth estimation problem is shown in Fig \ref{point matching principles}. The left panel shows the principle of visible-light stereo matching and depth estimation. The right panel shows the geometry of stereo X-ray tomographic imaging with two projections, which can also calculate the spatial information of a point in space from matched points in the two projections.

To increase the robustness of this approach to errors in the localisation of point features, we here instead use a learned feature matching and 3D mapping method. We train a 3D U-net to process the back-projected volumetric image of the detected feature maps to generate a 3D image containing the point and line features only. Formally, if $B_L(\cdot)$ and  $B_R(\cdot)$ are the filtered backprojection operators \cite{feldkamp1984practical} for the two projection images, then we train a mapping 
\begin{equation}
   \mathbf{y}_{vol} = g\left (B_L( \widehat{\mathbf{y}}_{mask}^L)+ B_R(\widehat{\mathbf{y}}_{mask}^R);\theta  \right ) \label{eq:1.2}
\end{equation}
from two extracted feature maps $\widehat{\mathbf{y}}_{mask}^{L}$ and $\widehat{\mathbf{y}}_{mask}^{R}$  to a 3D tomographic volume $\mathbf{y}_{vol}\in \mathbb{R}^{m\times n \times o}$.

\subsection{Proposed Framework}
\begin{figure*}
  \begin{center}
  \includegraphics[width=6.5in]{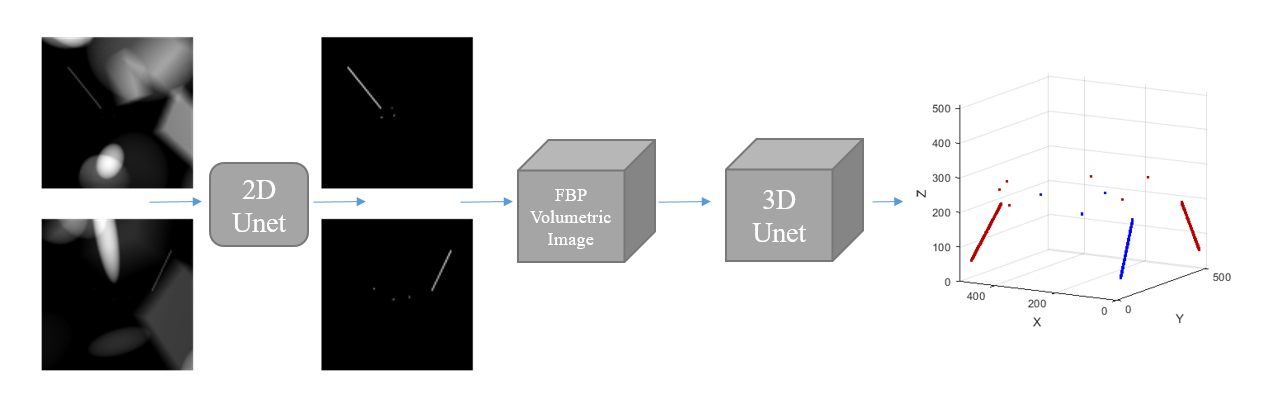}\\
  \caption{Overview of the proposed framework. The input is a pair of X-ray projection images which contain the X-ray projection of objects as well as the projection of line and point like features. Each projection is fed independently into the same 2D u-net to compute two different feature maps, where the background is removed. The two feature maps are then back-projected into a 3D volume using the FDK algorithm, which is then further processed using a 3D u-net to generate the 3D spatial feature maps (Then blue line and points are the 3D feature locations and the red is their projections on YZ/XZ planes). Note that for the 2D and 3D u-net in our framework, the input size is $256*256$ for 2D u-net, $512*512*512$ for 3D u-net, and the binary cross entropy loss function for the classification problem, different from the standard 2D/3D u-net.}\label{overview of framework}
  \end{center}
\end{figure*}

Our proposed approach is summarised in Fig. \ref{overview of framework}. We employ a  standard 2D u-net \cite{ronneberger2015u} as a feature detector $f(\cdot;\theta)$, and a similar standard 3D u-net \cite{cciccek20163d} to mode the mapping $g(\cdot,\cdot;\theta)$  to extracted projections to 3D feature maps. Both networks act as multi-label classification networks so we use a sigmoid output non-linearity together with the binary cross-entropy loss function.

%As shown in Fig. \ref{overview of framework}, we start from an unknown object with only its two X-ray projections via two X-ray sources. In order to extract point and line features from the raw projection, the 2D u-net is utilized as the binary classification network with the raw projection as input to extract feature mask that only contain the point and line features, but not the other objects that are visible in the raw projection. Feature maps are binary images, where 1 represents the location of a feature. 

% And then the prediction dot-multiplied with its corresponding raw X-ray image to obtain the extracted feature projection as Eq. \ref{eq:1.2}.  

%After extracting the feature from the pairs of X-ray projections, we use the FDK algorithm \cite{feldkamp1984practical} to map the features into a 3D volumetric image followed by a 3D u-net  to predict feature locations in 3D. As another simple binary classification network, we take the binary cross-entropy as our loss function for the training of the feature 3D volumetric mapping network.

%In the proposed framework, the feature detection network and feature volumetric reconstruction network use a 2D u-net and a 3D u-net to complete the 2D-to-2D-to-3D cross-domain transformation. 

% \begin{figure}
%   \begin{center}
%   \includegraphics[width=3.5in]{pdf/5.png}\\
%   \caption{2D unet structure.}\label{2D unet}
%   \end{center}
% \end{figure}

% \begin{figure}
%   \begin{center}
%   \includegraphics[width=3.5in]{pdf/6.png}\\
%   \caption{3D unet structure.}\label{3D unet}
%   \end{center}
% \end{figure}

\section{Datasets}
To train and test our approach, we use a synthetic and a real XCT dataset. The synthetic dataset uses randomly located point and line features superimposed over randomly placed polyhedra and spheres, each with varying attenuation. The real XCT data was generated during a proof of concept experiment to study the consolidation during carbon fibre tape layup, where thin copper wires were embedded in a carbon fibre tape to act as fiducial markers.  

\subsection{The Synthetic Dataset}
\begin{figure}
  \begin{center}
  \includegraphics[width=2.5in]{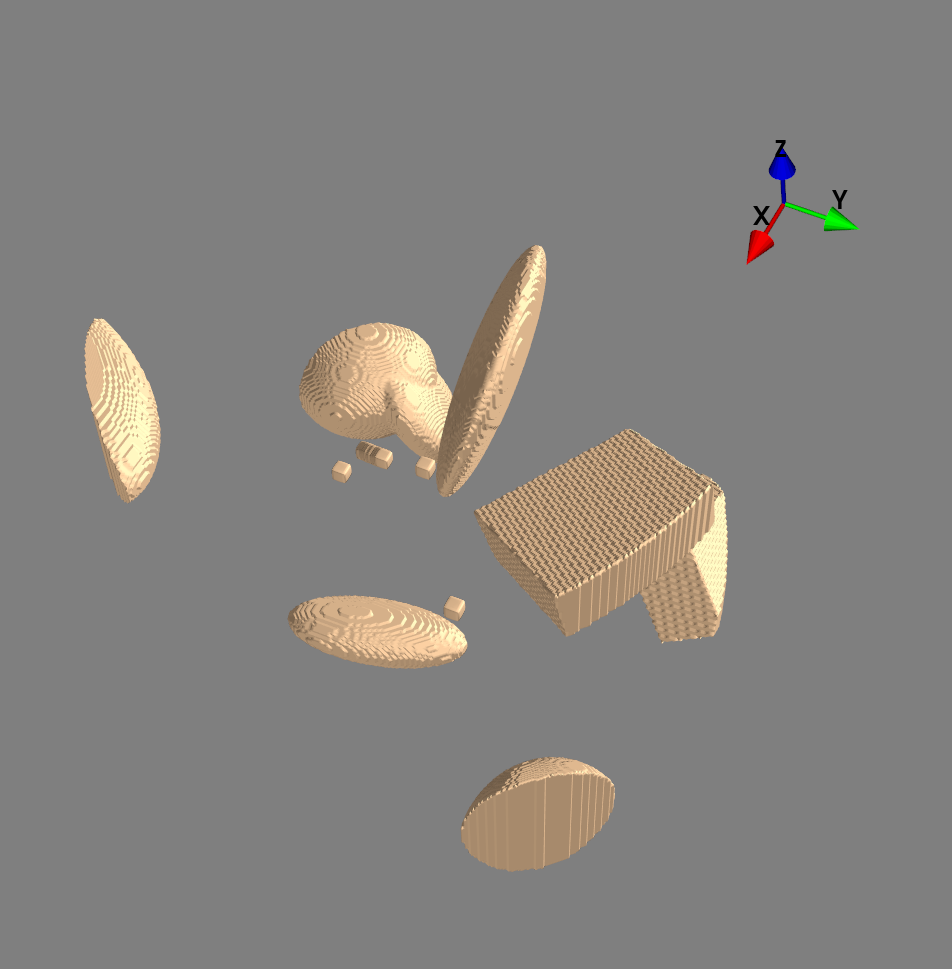}\\
  \caption{A rendering of an example of the 3D volumetric data showing polyhedra and ellipsoids as well as point and line features. }\label{one sample}
  \end{center}
\end{figure}

Our synthetic dataset consists of 100 3D images, generated by randomly generating 10 shapes (either polyhedra or ellipsoids), each with a random orientation, random dimensions and random attenuation values. To generate the volume, the attenuation of overlapping shapes is added to generate the attenuation in the overlapping region. The shapes are restricted to lie within the cylinder that is covered by the X-ray cone-beam by removing the parts of each object that lie outside the cylinder. 3 random points and one random line feature are then added. Point and line features are generated with 10 different attenuation values for each volume so that we have 1000 different volumes. Each volume is then projected twice using a cone-beam geometry generated using the Astra Toolbox \cite{van2015astra}, where the two projection directions are rotated by 90 degrees relative to each other.  A 3D rendering of an example is shown in Fig. \ref{one sample}. 

\begin{figure}[htbp]
\centering
\begin{tabular}{lll}
$\includegraphics[width=0.9in]{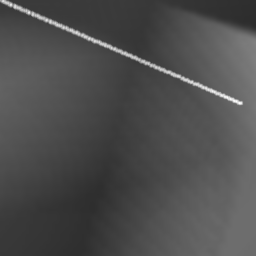}$ & $\includegraphics[width=0.9in]{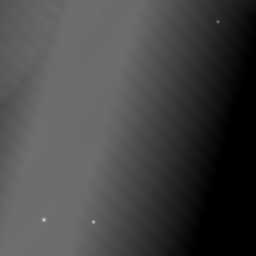}$ & $\includegraphics[width=0.9in]{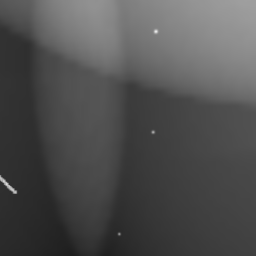}$  \\
$\includegraphics[width=0.9in]{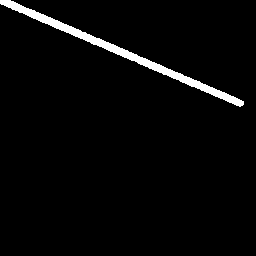}$ & $\includegraphics[width=0.9in]{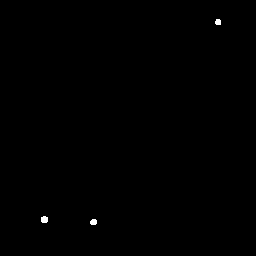}$ & $\includegraphics[width=0.9in]{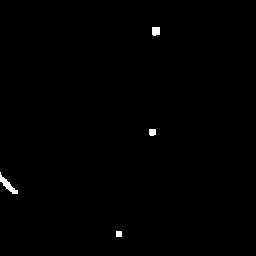}$ 
\end{tabular}\caption{Three sets of training samples (small blocks) for the feature detection method. The top images show the input data and the bottom images show the feature masks, which are used as the ground truth for feature detection.}\label{training samples}
\end{figure}

The original 3D volumes have $512*512*512$ voxels whilst the projections have $1024*1024$ pixels which can generate 144 overlapping blocks of $256*256$ pixels from each projection. We show three small blocks of randomly selected 2D training data pairs in Fig. \ref{training samples}. To make the synthetic data more realistic, for each image we have different intensities for the point and line features, gray values for point and line features are blurred with a Gaussian blur. All attenuation values are drawn from a uniform random distribution with values between 0 and 1.  

\subsection{The Carbon Fibre XCT dataset}

The real data was generated for the carbon fibre layup and consolidation experiment, where carbon fibre tape was deposited overlayup tool. The experiment was conducted in an X-ray scanner, and we collected x-ray projections at 60 different time points features the consolidation process. To track 3D deformation from stereo projections, two projection images were taken at each time point. 25 micrometer thick copper wires were embedded into the carbon-hand-annotated fiducial markers for 3D tracking of movement. The experiment was conducted with a bespoke test rig placed in a Nikon XTH225 X-ray tomography system. Images were acquired with a magnification of 8, giving a field of view of about 25mm so that each detector pixel is about 25 micrometers squared. To allow fast imaging and to reduce noise, the data from 2000 by 2000 pixel detector was binned into 1000 by 1000 pixels. The images are pre-processed by converting the measured X-ray intensity into attenuation values \cite{kak2001principles}.

\begin{figure}
\centering
\subfigure[Projection samples]{
\begin{minipage}[b]{0.43\linewidth}
\includegraphics[width=1\linewidth]{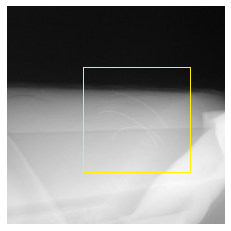}\vspace{4pt}
\includegraphics[width=1\linewidth]{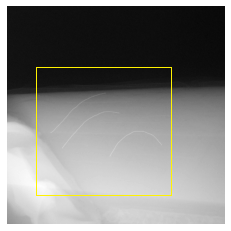}
\end{minipage}}
\subfigure[Ground truth samples]{
\begin{minipage}[b]{0.43\linewidth}
\includegraphics[width=1\linewidth]{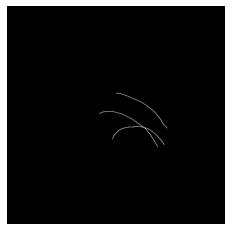}\vspace{4pt}
\includegraphics[width=1\linewidth]{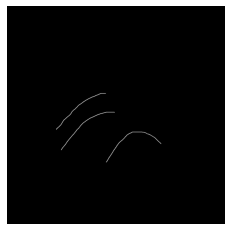}
\end{minipage}}
\caption{A pair of x-ray stereo projection images from a carbon fibre layup experiment. Two projections are shown on the left and hand annotated images that show the feature location are shown on the right. We limited the evaluation of the confusion matrix into the yellow box to reduce the occurrences of non-feature pixels to better present evaluation results.%The yellow box is the region of interest to facilitate the analysis of the accuracy on test projections. 
}\label{2 large image samples}
\end{figure}

The dataset thus consists of 60 pairs of projection images. The lines were annotated in each projection by hand. Two different projection images are shown on the left in Fig. \ref{2 large image samples}, where we also show the hand annotated locations of three linear features on the right. As the copper wires produced in the original projections are extremely faint, we also generated augmented datasets, where we changed the attenuation value along these features, by either doubling or halving the values.

\section{Experimental evaluation}
We evaluate the feature detection and the 3D feature mapping steps independently. 

\subsection{General training and evaluation approach}

The 2D classification network for feature detection is trained using the projections as inputs and the binary images showing point locations as output. As for the 3D volumetric reconstruction experiment, based on the Eq. \ref{eq:1.2}, the inputs are the filtered back-projection volume images generated from the feature maps detected with the feaqture detection network.
% \begin{figure}
%   \begin{center}
%   \includegraphics[width=3.5in]{pdf/7.png}\\
%   \caption{}\label{a sample on feature reconstruction}
%   \end{center}
% \end{figure}
Both networks are implemented using TensorFlow 2.0 and optimised using an Nvidia Titan XP graphics card. We use the Adam optimiser with a learning rate of $10^{-4}$ run for 100 epochs. 

%And to evaluate the performance of feature volumetric reconstruction, we compare the center positions of point features, the endpoint positions of line features with the absolute pixels difference, and intensity histograms between the reconstruction and the ground truth.

\subsection{Feature Detection}

Our first feature detection experiment uses the synthetic data-set. We split the data into test and train sets so that we use 95 of the volumes for training and the projections from the remaining 5 volumes for testing. Note that for each of these volumes, we have 10 different feature intensities for each volumes and two projections from each, so that we have 1900 projections for training and 100 independently generated projections for testing.

\begin{figure}[htbp]
\centering
\begin{tabular}{ccc}
Input & Output & Truth  \\
$\includegraphics[width=0.9in]{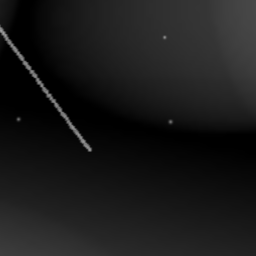}$    & $\includegraphics[width=0.9in]{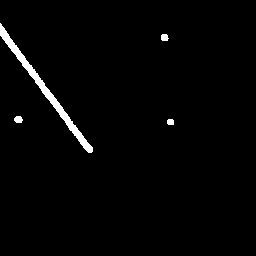}$     & $\includegraphics[width=0.9in]{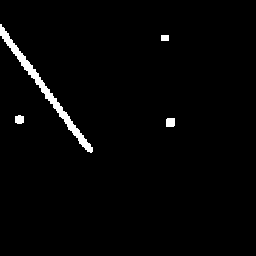}$     \\
$\includegraphics[width=0.9in]{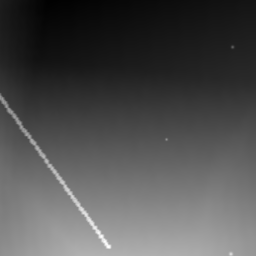}$    & $\includegraphics[width=0.9in]{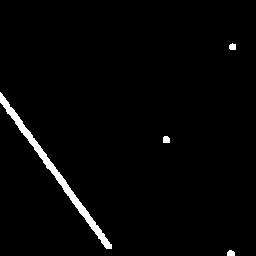}$     & $\includegraphics[width=0.9in]{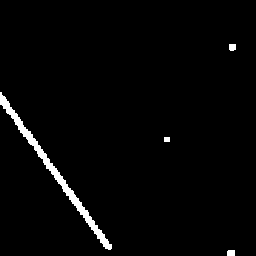}$     \\
$\includegraphics[width=0.9in]{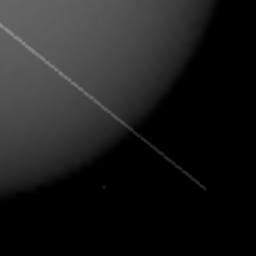}$    & $\includegraphics[width=0.9in]{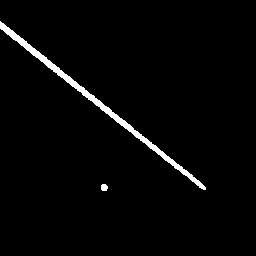}$     & $\includegraphics[width=0.9in]{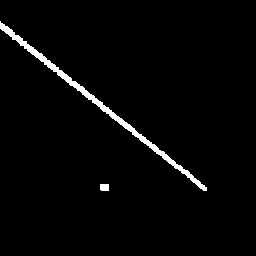}$    
\end{tabular}\caption{Three test samples with features at different gray-levels. The first column shows the 2D projection, the second column is the estimated feature locations and the last column is the ground truth.}\label{Feature result s-s}
\end{figure}

\begin{figure}
  \begin{center}
  \includegraphics[width=3.5in]{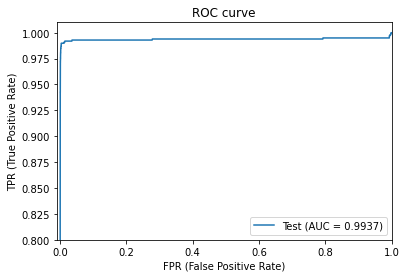}\\
  \caption{ROC curve for our test data for the synthetic data-set. The AUC is 0.9937.}\label{ROC}
  \end{center}
\end{figure}

Once trained, the performance of the method on 3 of the test samples (small blocks) is demonstrated in Fig. \ref{Feature result s-s}, where we show three examples, with predicted feature locations on the left, ground truth locations in the centre and the original projections on the right. To evaluate the performance numerically, we show the normalised confusion matrix in Table. \ref{synthetic test data CM}. We also show the ROC curve (receiver operating characteristic curve) in Fig. \ref{ROC}. Our test performance curve almost reaches the effect of the perfect model, indicated from the area under the curve (AUC) is close to 1.

% \begin{figure}
%   \begin{center}
%   \includegraphics[width=3.5in]{pdf/10.png}\\
%   \caption{Confusion matrix for the test data. In our confusion matrix, we define 0 as negative, and 1 as positive.}\label{CM}
%   \end{center}
% \end{figure}

\begin{table*}
\centering
\begin{tabular}{|c|c|c|c|l|} 
\hline
\multicolumn{2}{|c|}{\multirow{2}{*}{Confusion matrix}}             & \multicolumn{2}{c|}{\begin{tabular}[c]{@{}c@{}}\\Predicted condition\\\end{tabular}}                                                        &                                             \\ 
\cline{3-4}
\multicolumn{2}{|c|}{}                                              & \begin{tabular}[c]{@{}c@{}}\\Positive\\\end{tabular}                   & Negative                                                           &                                             \\ 
\hline
\multirow{2}{*}{Actual condition} & Positive              & \begin{tabular}[c]{@{}c@{}}\\TP = 719\\\end{tabular}                 & \begin{tabular}[c]{@{}c@{}}\\FN = 52\\\end{tabular}              & \multicolumn{1}{c|}{TPR=TP/(TP+FN)=0.933}   \\ 
\cline{2-5}
                                            & Negative              & \begin{tabular}[c]{@{}c@{}}\\FP = 37\\\end{tabular}                  & TN = 64728                                                      & \multicolumn{1}{c|}{FPR=FP/(FP+TN)=0.0006}  \\ 
\hline
\multicolumn{1}{|l|}{}                      & \multicolumn{1}{l|}{} & \begin{tabular}[c]{@{}c@{}}\\PPV = \\TP/(TP+FP) = 0.951\\\end{tabular} & \begin{tabular}[c]{@{}c@{}}FOR =\\FN/(FN+TN) = 0.0008\end{tabular} &                                             \\
\hline
\end{tabular}\caption{Confusion matrix on synthetic test data.}\label{synthetic test data CM}
\end{table*}

% \begin{table}
% \centering
% \begin{tabular}{|c|c|c|c|} 
% \hline
% \multicolumn{2}{|c|}{\multirow{2}{*}{Confusion Matrix}} & \multicolumn{2}{c|}{True}  \\ 
% \cline{3-4}
% \multicolumn{2}{|c|}{}                                  & Negative & Positive        \\ 
% \hline
% \multirow{2}{*}{Predict} & Negative                     & TN=64728 & FN=37           \\ 
% \cline{2-4}
%                          & Positive                     & FP=52    & TP=719          \\
% \hline
% \end{tabular}\caption{Confusion matrix. True Positive (TP), False Negative (FN), False Positive (FP) and True Negative (TN) are the most basic index of the confusion matrix.}\label{1st index}
% \end{table}

\begin{figure}
  \begin{center}
  \includegraphics[width=3.5in]{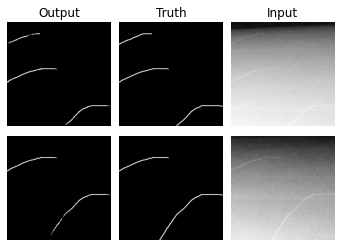}\\
  \caption{Results on two test samples. The first column is the predicted feature location, the second column is the ground truth feature location, and the last column is the original projection image.}\label{cf2 result samples}
  \end{center}
\end{figure}

\begin{figure}
  \begin{center}
  \includegraphics[width=3.5in]{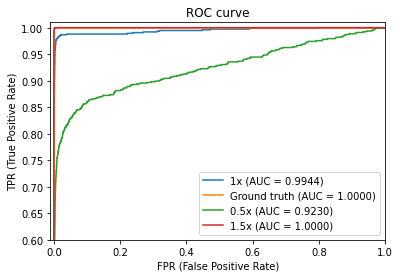}\\
  \caption{ROC curve for the real data but with varying feature attenuation. The red curve shows features that are 1.5 times stronger than those in the original data, the blue curve is the original feature strength and the green curve is the performance if features have half the attenuation. The stronger features lead to perfect recovery, whilst the AUC for the less strong features drops to 0.9230.}\label{ROC comparsion}
  \end{center}
\end{figure}

As true X-ray data have larger projection images but fewer training examples, we train the network on image blocks. From the 120 projection images, we thus generate 11520 overlapping blocks of $256*256$ pixels and use these as our training set.
We split the data into test and train sets by taking one projection direction as the training example and the comparison projection direction as the test sample. Example results are shown in  Fig. \ref{cf2 result samples}, where we again show predicted feature locations (left), ground truth (middle) as well as projections (right). The confusion matrix is shown in Table. \ref{real test data on CM} and the ROC blue curve in Fig. \ref{ROC comparsion}. 

% \begin{figure}
%   \begin{center}
%   \includegraphics[width=3.5in]{pdf/13.png}\\
%   \caption{Confusion matrix on the average test data.}\label{CM2}
%   \end{center}
% \end{figure}

\begin{table*}
\centering
\begin{tabular}{|c|c|c|c|l|} 
\hline
\multicolumn{2}{|c|}{\multirow{2}{*}{Confusion matrix}}             & \multicolumn{2}{c|}{\begin{tabular}[c]{@{}c@{}}\\Predicted condition\\\end{tabular}}                                                        &                                             \\ 
\cline{3-4}
\multicolumn{2}{|c|}{}                                              & \begin{tabular}[c]{@{}c@{}}\\Positive\\\end{tabular}                   & Negative                                                           &                                             \\ 
\hline
\multirow{2}{*}{Actual condition} & Positive              & \begin{tabular}[c]{@{}c@{}}\\TP = 1214\\\end{tabular}                 & \begin{tabular}[c]{@{}c@{}}\\FN = 89\\\end{tabular}              & \multicolumn{1}{c|}{TPR=TP/(TP+FN)=0.932}   \\ 
\cline{2-5}
                                            & Negative              & \begin{tabular}[c]{@{}c@{}}\\FP = 264\\\end{tabular}                  & TN = 129505                                                      & \multicolumn{1}{c|}{FPR=FP/(FP+TN)=0.002}  \\ 
\hline
\multicolumn{1}{|l|}{}                      & \multicolumn{1}{l|}{} & \begin{tabular}[c]{@{}c@{}}\\PPV = \\TP/(TP+FP) = 0.821\\\end{tabular} & \begin{tabular}[c]{@{}c@{}}FOR =\\FN/(FN+TN) = 0.0001\end{tabular} &                                             \\
\hline
\end{tabular}\caption{Confusion matrix on real test data.}\label{real test data on CM}
\end{table*}
% \begin{table}
% \centering
% \begin{tabular}{|c|c|c|c|} 
% \hline
% \multicolumn{2}{|c|}{\multirow{2}{*}{Confusion Matrix}} & \multicolumn{2}{c|}{True}  \\ 
% \cline{3-4}
% \multicolumn{2}{|c|}{}                                  & Negative & Positive        \\ 
% \hline
% \multirow{2}{*}{Predict} & Negative                     & TN=129505 & FN=264           \\ 
% \cline{2-4}
%                          & Positive                     & FP=89    & TP=1214          \\
% \hline
% \end{tabular}\caption{Confusion matrix index.}\label{1st index2}
% \end{table}

% \begin{figure}
%   \begin{center}
%   \includegraphics[width=3.5in]{pdf/14.png}\\
%   \caption{ROC curve on the real x-ray data. For better visualization, we  only show part of the true positive rate axis. %The blue curve belongs to ROI in test data, the orange one belongs to ground truth, and 
%  The AUC of the test data is 0.9944.}\label{ROC2}
%   \end{center}
% \end{figure}

\begin{figure}
\centering
\begin{tabular}{rccc}
\begin{sideways}Intensity\end{sideways} & 0.5x  & 1x    & 1.5x   \\
\begin{sideways}Input\end{sideways}     & $\includegraphics[width=0.9in]{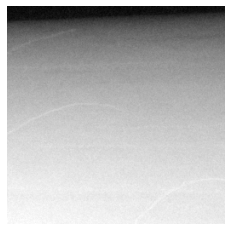}$ & $\includegraphics[width=0.9in]{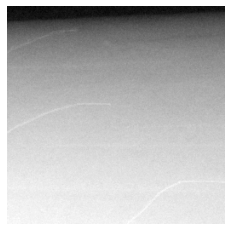}$ & $\includegraphics[width=0.9in]{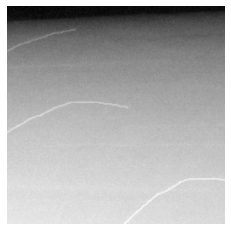}$  \\
\begin{sideways}Output\end{sideways}    & $\includegraphics[width=0.9in]{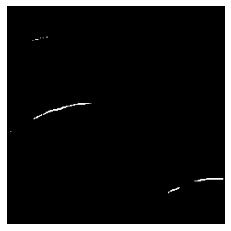}$ & $\includegraphics[width=0.9in]{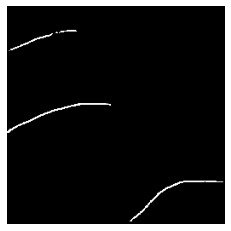}$ & $\includegraphics[width=0.9in]{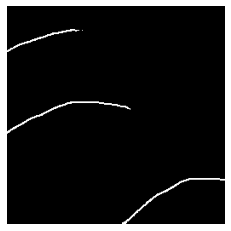}$  \\
                                        & (a)   & (b)   & (c)   
\end{tabular}\caption{A comparison of the model's performance with features of different strengths. From left to right, we show features that are 0.5 time, 1 times and 1.5 times as strongly attenuating relative to the features in the original data. The top row shows the same block from one projection image but with different feature attenuation and the bottom row shows the detected features.}\label{intensity comparsion}
\end{figure}

To further explore the influence of the intensity of the feature relative to the background attenuation in the projection image, we compare the performance of the model by training the modified dataset if we have features that are halve or 1.5 times as strongly attenuating. Example images with differently strong features are shown in Fig. \ref{intensity comparsion}. Here we use 0.5 times to represent weak intensity, 1 times for normal intensity and 1.5 times for strong intensity. Classification performance is shown in Fig. \ref{ROC comparsion}, where we show the ROC curves for different feature intensities. As long as the features are not too weak, our model correctly detects the features.

\subsection{Experiments of Feature Volumetric Reconstruction}

To evaluate our approach to feature matching and 3D location estimation, we use the synthetic data-set only, as we do not have the ground truth for the real dataset. We train the 3D U-net on the estimated feature maps from the training dataset, using the ground truth 3D location of feature as the target. We then evaluate the method using the test data. 

\begin{figure}
\centering
\begin{tabular}{cc}
First view & Second view  \\
$\includegraphics[width=1.6in]{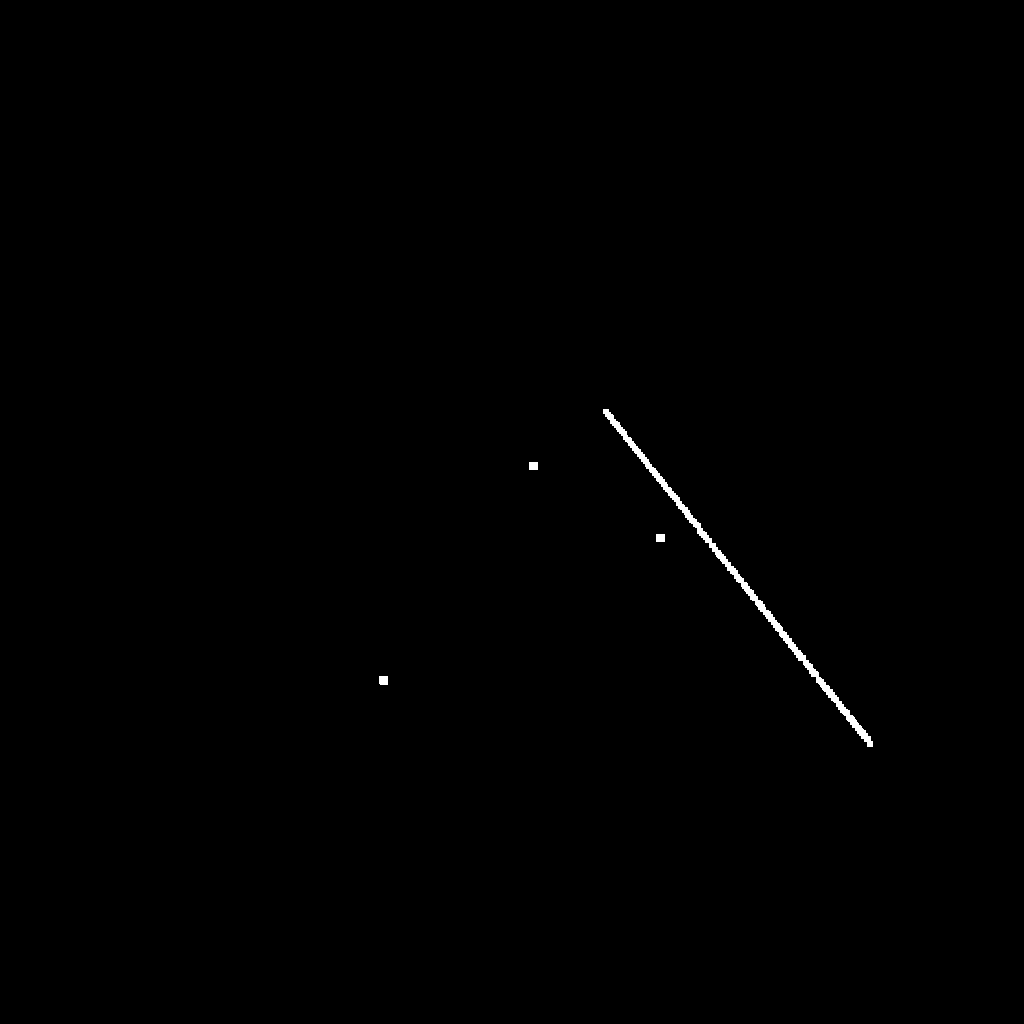}$       & $\includegraphics[width=1.6in]{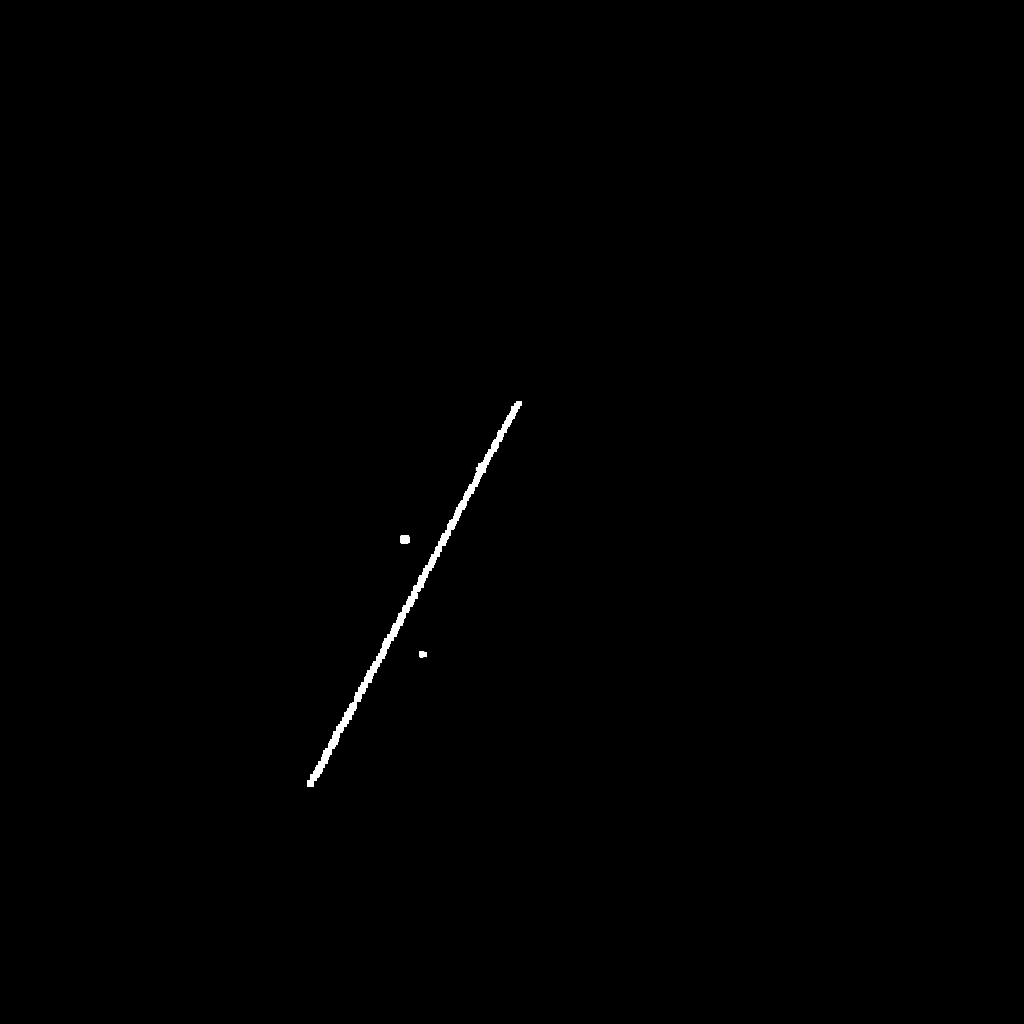}$     \end{tabular}\caption{A pair of projections from a synthetic test sample. The first view shows all three point features as well as the line feature, whilst in the second view, a point feature overlaps with the line feature and is thus not visible.}\label{occulsion case}
\end{figure}

\begin{figure}
\centering
\begin{tabular}{cc}
3D mapping & Ground truth  \\
$\includegraphics[width=1.6in]{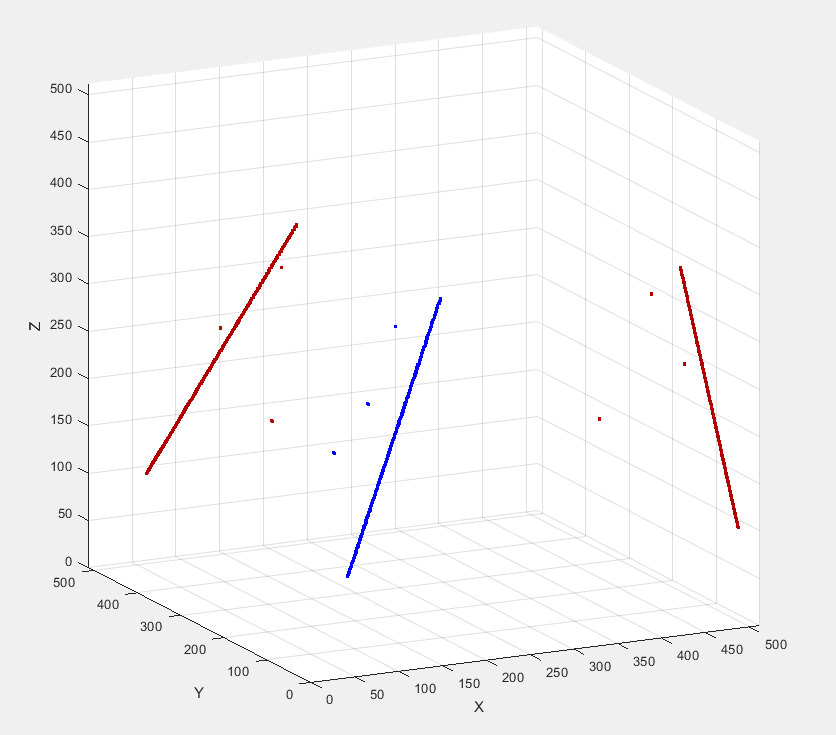}$       & $\includegraphics[width=1.6in]{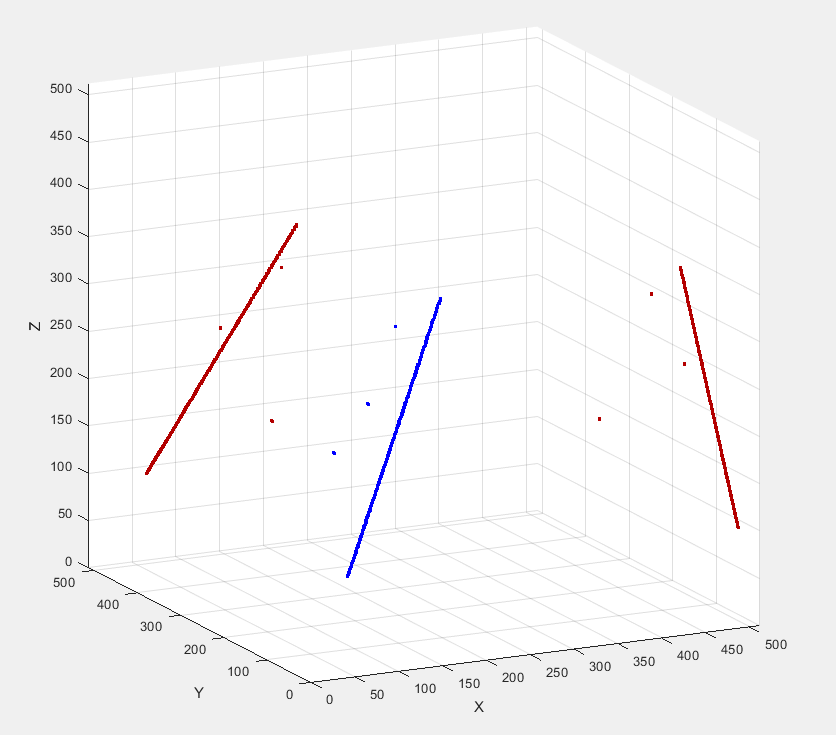}$     \end{tabular}\caption{Rendering of the 3D location of the features reconstructed from the estimated feature positions in the projections as shown in Fig. \ref{occulsion case}, showing that the method still reconstructs the features, even though they are not all visible in both projection views.}\label{mapping result and gt}
\end{figure}

Fig. \ref{occulsion case} shows an example where the geometry does not allow us to find unique 3D mappings, as in the second projection image, a point feature has been occluded by the line feature, so that the point feature from the first view can only be match with its corresponded point feature in the other view if we assume that two features overlap in one image. We find that even for this difficult case, the neural network can handle this uncertainty, with the rendering of the true and estimated 3D feature locations for this example shown in Fig. \ref{mapping result and gt}. We list 10 of our test samples with their average confusion matrix at Table. \ref{3D mapping CM} and numerically evaluate the performance of this approach compare the absolute difference between the centre position of point features and line feature end point centers for the 10 test samples. We found that the average absolute difference is lower than 1.5 pixels.  

\begin{table*}
\centering
\begin{tabular}{|c|c|c|c|l|} 
\hline
\multicolumn{2}{|c|}{\multirow{2}{*}{Confusion matrix}}             & \multicolumn{2}{c|}{\begin{tabular}[c]{@{}c@{}}\\Predicted condition\\\end{tabular}}                                                        &                                             \\ 
\cline{3-4}
\multicolumn{2}{|c|}{}                                              & \begin{tabular}[c]{@{}c@{}}\\Positive\\\end{tabular}                   & Negative                                                           &                                             \\ 
\hline
\multirow{2}{*}{Actual condition} & Positive              & \begin{tabular}[c]{@{}c@{}}\\TP = 12206\\\end{tabular}                 & \begin{tabular}[c]{@{}c@{}}\\FN = 2273\\\end{tabular}              & \multicolumn{1}{c|}{TPR=TP/(TP+FN)=0.843}   \\ 
\cline{2-5}
                                            & Negative              & \begin{tabular}[c]{@{}c@{}}\\FP = 2119\\\end{tabular}                  & TN = \num{1.34e8}                                                      & \multicolumn{1}{c|}{FPR=FP/(FP+TN)=\num{1e-5}}  \\ 
\hline
\multicolumn{1}{|l|}{}                      & \multicolumn{1}{l|}{} & \begin{tabular}[c]{@{}c@{}}\\PPV = \\TP/(TP+FP) = 0.852\\\end{tabular} & \begin{tabular}[c]{@{}c@{}}FOR =\\FN/(FN+TN) = \num{1e-5}\end{tabular} &                                             \\
\hline
\end{tabular}\caption{Confusion matrix for 3D mapping result.}\label{3D mapping CM}
\end{table*}

\begin{figure}
  \begin{center}
  \includegraphics[width=3.5in]{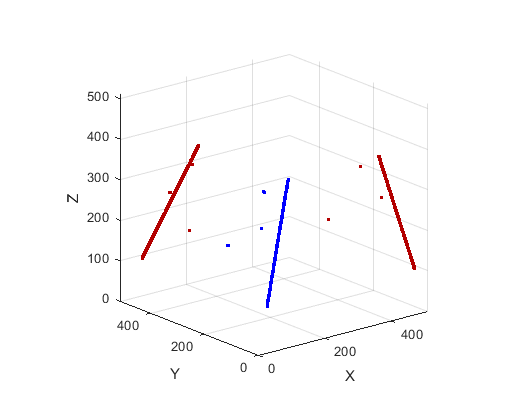}\\
  \caption{The feature mapped from two feature detected projections into 3D space by epipolar geometric method with a replaced projection on second view due to a point feature is occluded by the line feature.}\label{epipolar geometric 3D mapping}
  \end{center}
\end{figure}

To demonstrate the improved performance of our neural network approach, we also match the features of Fig. \ref{occulsion case} and map them into 3D space via the epipolar geometric information as our reference. In this epipolar geometric method, the intersection point in 3D space of the back-projection generated by two different single views is the position of the feature point in 3D space, which is mathematically verifiable and accurate. However, considering the voxel and pixel size in 3D space and 2D projections in the digital image, there will inevitably be errors when selecting the center of the feature point. As shown in Fig \ref{epipolar geometric 3D mapping}, the absolute pixels error for the three points feature and two points feature at the end of line features is 1. Comparing Fig \ref{epipolar geometric 3D mapping} with Fig. \ref{mapping result and gt}, our neural network method performs as well as the epipolar geometric method but crucially also works in some of the cases where the epipolar method fails due to point occlusion.

\section{Discussion and conclusions}
In this paper, we introduce the concept of stereo X-ray tomography, where we use two (or possibly 3) X-ray projection images to estimate spatial locations of features in 3D space. Whilst we are not able to reconstruct arbitrary objects in 3D without additional prior information, we show that it is possible to reconstruct the location of the point and line features in 3D. This can have many applications in tomographic imaging, especially when we are unable to collect full tomographic projections, which is useful when mapping dynamic processes that are too fast for full tomographic acquisition. For these imaging systems, setups with two (or three) X-ray source and detector systems can be envisaged that inspect an object at roughly 90-degree angles. In this setting, the main challenge is the identification and matching of points in the individual projections. Methods used in stereo vision, which typically rely on feature matching methods that match entire pixel neighbourhoods, do not work in transmission tomography. Instead, we propose the use of a learned feature detector together with a feature matching method that exploits epipolar geometry constraints. We have shown the robustness of the feature detection method and could demonstrate that, for problems with few features where unique matching is possible, a simple 3D U-net can map back-projected feature maps into 3D locations. It should be stressed that for two projection images (binocular Stereo CT), feature matching can become a challenge when using the epipolar constraint if we have larger numbers of features as matching is not uniquely possible if two features lie on the same epipolar plane in 3D. In this case, matching is however much more likely if we have 3 projections (triocular Stereo CT), such that none of the equipolar planes are parallel.

\ifCLASSOPTIONcaptionsoff
  \newpage
\fi



%\begin{thebibliography}{1}


\bibliographystyle{IEEEtran}
\bibliography{IEEEabrv,Bibliography}

\vfill


\end{document}